# Three Metrics for Measuring User Engagement with Online Media and a YouTube Case Study

Lassi A. Liikkanen

Helsinki Institute for Information Technology HIIT, Aalto University

P.O.Box 15600, FI-00076 AALTO, Finland

e-mail address: lassi.liikkanen@hiit.fi, Twitter: @lassial

## ABSTRACT

This technical report discusses three metrics of user engagement with online media. They are Commenting frequency, Voting frequency, and Voting balance. These relative figures can be derived from established, basic statistics available for many services, prominently YouTube. The paper includes case a study of popular YouTube videos to illustrate the characteristics and usefulness of the measures. The study documents the range of observed values and their relationships. The empirical sample shows the three measures to be only moderately correlated with the original statistics despite the common numerators and denominators. The paper concludes by discussing future applications and the needs of the quantification of user interaction with new media services.

## ACM Classification Keywords

H.5.2. Information interfaces and presentation (e.g., HCI): Evalution/Methodology.

## INTRODUCTION

Online media services, such as YouTube, provide new ways for users to engage and react to media content. For instance, YouTube visibly promotes voting (thumbs up and thumbs down) and commenting of the videos. Together with the total number of views, this produces the basic video statistics. Additionally the service collects many statistics of user behavior invisibly without consent. For instance, YouTube records retention for each video and watching session. This and some other statistics are by default only available to the video and content owners.

YouTube is currently the third most used website globally[1]. Thus it comes as no surprise that the public statistics, foremost video views, of YouTube have attracted much attention also from news media[2]. However, given the fact that YouTube has never publicly announced what counts as 'a view', numbers such as these should be considered with caution. Similarly, the aggregate numbers of user comments or votes, which are technically straightforward to define, can hardly be considered as very usable in their own; without a reference point. This is because people cannot easily interpret or compare these kind of substantial numbers without prior knowledge.

The previous studies of user behavior and aggregate use characteristics with online video (YouTube) have typically resorted just to using the basic public statistics [1, 2, 4, 5]. Many of these reports report the total or average numbers Views or Comments at a given time. However, the problem of this approach is that these absolute numbers get quickly out of date. User engagement with the content can continue as long the media content remains accessible and by definition this means the figures can only grow bigger. For instance, an artist may become unfashionable due to bad publicity causing users to react more negatively than they initially did. If we only look at absolute figures this change could be easily dismissed. I believe more information can be gained if we have *relative* metrics instead of absolute indexes for audience interaction.

In this paper, I propose three aggregate level statistics that can help scholars and journalists to interpret the absolute figures of audience engagement. The three metrics measure *Commenting frequency*, *Voting frequency*, and *Voting balance*. They are operationalized in measures called *Comments per thousand Impressions* (CpkI), *Votes per thousand Impressions* (VpkI), and *Dislike Proportion* (DisP). They are applicable to services counting impressions (Views or Plays), allow commenting, and include binary voting (+/-, thumb up/thumb down) option. Importantly, these figures can be derived from public statistics, unlike private YouTube Analytics statistics.

The metrics are intended to allow comparisons between videos that attract different volumes of attention. Variations of VpkI and DisP have both appeared once before in the literature (VpkI in [6], DisP in [3] and an equivalent metric in [1]), but it seems timely to present them together for the first time and unify the terminology so it can be consistently used in upcoming studies.

---

Additionally, I will present example data drawn from popular YouTube videos to illustrate the range of the metrics and assess their interdependence. The finding is that the three measures can all provide new insights about online media. As this research establishes some reference levels for these metrics, researchers can in future more easily compare their findings of online media use.

## NEW METRICS

The three metrics are derived from four existing metrics: the total number of impressions ($N_i$), the number of positive votes ($N_{v+}$), the number of negative votes ($N_{v-}$), and the number of comments ($N_c$). Based on the availability of these numbers, we can calculate the Comments per thousand impressions (CpkI). It tells us how many impressions produce one comment on average:

$$CpkI = \frac{N_C \times 1000}{N_i} \qquad \text{(Eq. 1)}$$

In a similar fashion we derive Votes per thousand Impression (VpkI), which measures the frequency of voting. VpkI is only defined when $N_{v-} > 0$:

$$VpkI = \frac{(N_{v+} + N_{v-}) \times 1000}{N_i} \qquad \text{(Eq. 2)}$$

Dislike proportion (DisP) presents the share of negative votes. The calculation of DisP is performed as follows:

$$DisP = \frac{N_{v-}}{(N_{v+} + N_{v-})} \qquad \text{(Eq. 3)}$$

DisP ranges from 0-1. We have thus defined CpkI, VpkI, and DisP in three equations (1, 2, and 3).

## CASE STUDY: YOUTUBE VIDEOS

The three metrics can be immediately applied to YouTube videos. As of late 2013, YouTube displays the necessary data in association with each video, total video views equaling to impressions. The relevant measurements can also be conveniently retrieved using the YouTube API and automatically computed with the formulae.

To illustrate the range and type of the measures, we calculated and analyzed CpkI, VpkI, and DisP for trending YouTube videos. 100 currently popular videos (as determined by YouTube algorithms) and their statistics were retrieved using API version 3 in December 2013. In conflict with the documentation, the API returned only 50 unique IDs at a time, so sampling was done over a span of a week on three occasions to retrieve 106 unique video IDs. The hundred videos with the greatest number of Views and allowed commenting were chosen for the sample reported here. The Appendix 1 contains the list of the sampled video IDs.

## Descriptive Statistics

The basic statistics for the sample were summarized first. The resulting descriptive statistics are shown in Table 1.

|  | Views | Comments | Votes |
|---|---|---|---|
| **N** | 100 | 100 | 100 |
| **Average** | 2,456,693 | 3,526.0 | 25,217 |
| **Std. Dev.** | 527,0278 | 13,217 | 82,526 |
| **Minimum** | 7105 | 9 | 75 |
| **Maximum** | 36,285,216 | 107,059 | 642,878 |

**Table 1.** Descriptive statistics for the trending video sample of Views, Comments, and Votes.

The number of views, comments, votes have a considerable range of variation. The included videos have on average over two million views. The number of comments are not independent of views. The videos represent 14 different categories. The list of their frequencies in Appendix 2. The most prevalent category was Entertainment with 24 videos, followed by Tech (15 videos) and Sports (11 videos).

We then calculated the new metrics, as presented in Table 2. It shows that the three measures have an extensive range of variation. In addition to raw figures, we also binned the three variables to illustrate mode and median values. The binned distributions have a single peak in the lower end of the values as shown in the Figure 1 (next p).

|  | CpkI | VpkI | DisR |
|---|---|---|---|
| **Valid N** | 100 | 100 | 100 |
| **Mean** | 2.687 | 10.497 | 10.44% |
| **Std. Dev.** | 10.995 | 10.467 | 13.78% |
| **Bin Mode** | .6-1.0 | 2.0-4.0 | ≤ 4% |
| **Skewness** | 9.421 | 2.570 | 3.339 |
| **Kurtosis** | 91.976 | 8.434 | 13.961 |
| **Minimum** | 0.195 | 1.285 | 0.75% |
| **Maximum** | 109.354 | 63.723 | 88.27% |

**Table 2.** Descriptive statistics for CpkI, VpkI and DisP in the case sample.

## Interdependence of the Metrics

The bivariate correlations for all basic statistics and the three new measured were calculated and summarized for Table 3. Looking at the whole data, the table shows that CPKI and VpkI metrics are weakly correlated (r=.194, p>.05), but DisP mildly correlates with both VpkI (r=-

.208, p=.038). Importantly, all three variables are nearly independent of the basic statistics – with the exception of DisP which correlates the number of negative votes (Dislikes). This is expected due to its definition.

| | CpkI | VpkI | DisR | Views | V+ | V- | Comm |
|---|---|---|---|---|---|---|---|
| **VpkI** | 0.19 | 1 | | | | | |
| **DisR** | .016 | -.21* | 1 | | | | |
| **Views** | -.05 | -.01 | -.04 | 1 | | | |
| **Votes+** | -.02 | .16 | -.09 | **.90**\*\* | 1 | | |
| **Votes-** | .00 | .06 | .33** | **.50**\*\* | **.45**\*\* | 1 | |
| **Comments** | .04 | 0.12 | .014 | **.86**\*\* | **.92**\*\* | **.72**\*\* | 1 |
| **Votes (sum)** | -.02 | .16 | -.05 | **.90**\*\* | **.99**\*\* | **.54**\*\* | **.95**\*\* |

**Table 3.** Variable dependencies measured by Pearson R. Two asterisks mark significance levels p<.001 (N=100), bold typeface shows at least moderate R (>.4).

However, a more detailed inspection of the correlations against different magnitudes of Impressions shows that the reality is not so straightforward. In a quartile split of the data on Impressions, there emerge notable correlations beyond the first quartile (videos with > 300,000 Views), which are summarized in Table 4.

| | CpkI | VpkI | DisR | Views | V+ | V- | Comm |
|---|---|---|---|---|---|---|---|
| **VpkI** | **.723**\*\* | 1 | | | | | |
| **DisR** | -.075 | -.242* | 1 | | | | |
| **Views** | .251* | .051 | -.025 | 1 | | | |
| **Votes+** | **.375**\*\* | .234* | -0.1 | **.897**\*\* | 1 | | |
| **Votes-** | **.48**\*\* | .11 | **.409**\*\* | **.492**\*\* | **.443**\*\* | 1 | |
| **Comments** | **.486**\*\* | 0.176 | .027 | **.869**\*\* | **.918**\*\* | **.717**\*\* | 1 |
| **Votes (sum)** | **.408**\*\* | .233* | -.049 | **.903**\*\* | **.995**\*\* | **.53**\*\* | **.948**\*\* |

**Table 4.** Pearson R correlations for variables based on the data from three highest quartiles of Impressions (N=75).

Among the videos with more impressions, there is a moderately strong correlation between VpkI and CpkI, but DisP remains only weakly correlated. Commenting frequency CpkI is also moderately correlated with the number of negative votes, total number of comments and total number of votes, showing more troublesome aspects.

Correlations between the basic statistics were notable. As expected, the number of comments and votes increases quite linearly with the total number of views. However, the number of 'thumbs down' had a weaker, but still a positive connection. Voting and commenting were also very strongly correlated, hinting that the same content evokes users to both vote and comment.

**DISCUSSION**

This paper has suggested some new measures of interaction suitable for measuring audience engagement with online media. Because they are based on YouTube public statistics, these can be readily applied for media research. I have also illustrated the typical values and the relationship of the variables in a sample of popular YouTube videos. These demonstrate that the proposed are informative on the phenomenon. First, it showed that the three measures bring up dimensions of data beyond the original statistics – showing that despite the strong correlation of votes and comments and the views there is interesting variation.

The example data showed that among trending videos, users are typically quite active to vote and comment. However, due to the considerable variation, on average, there is one vote for every two hundred views and one comment for 1600 views. This makes voting clearly the more sensitive measurement, from which one might predict commenting when a video is yet to become popular. The trending videos were also generally liked. The average Dislike ratio was below 10%, indicating that people generally react positively to the videos. Both VpkI and DisP appear promising for future use as they appear quite independent and informational rich in comparison to basic statistics they were derived from

I hope that in future these measures will be further explored for their usefulness in online media research. These metrics should be utilized with some precautions. First the interdependencies must be kept in mind if the metrics are subjected to an analysis of variance or other test of statistical distribution. Second, the user interpretation of the underlying basic statistics may change. For instance, a design change in YouTube may influence how eagerly consumers comment on the videos or new layout of the player page may cause people to react differently and vote differently on the same content. This should be remembered when comparing the metrics across points in time; i.e. the observed change in preference may not be solely attributable to a change in consumer perception of the media object, but the whole viewing experience.

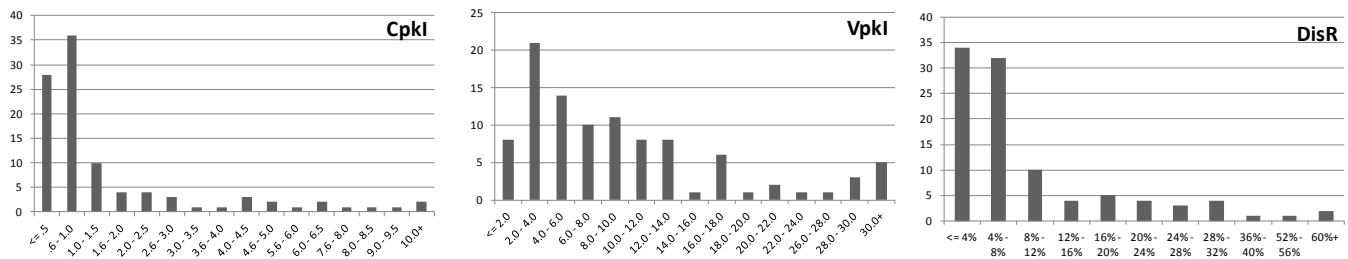

**Figure 1. Bin frequency distributions for CPKI, VpkI, and DisP.**

**APPENDIX 1**

List of sampled video IDs in alphabetical order separated by semicolons and a space. Video page on YouTube website can be accessed at http://youtu.be/ID

```
ZUrsZtcTaX4; _CzBlSXgzqI;     EpnERlsfBFc; FC5FbmsH4fw;     r4_-yXiRcjM; rfPnCPGg4cU;
06GhXB2_XNE; 1gNZ5qgEyWM;     FF_m6HBPufA; GKLZ5jSxPIc;     RrGPtCdItBw; S6vIuSQPlzk;
litrSlRH90A; 1pYdZlLOVK0;     GkSSzx6wTFk; Grp5-bvMo7U;     SctCxERhfFI; sdCBqI3Jn3Y;
1WBgiTqDvjc; 2hCdtUxnOG8;     GVCzdpagXOQ; gyBhdpyhcCw;     smnxo-qwzdA; sNPp74zh81M;
2hfre4eBCY8; 3N_o2JqVdwA;     H7jtC8vjXw8; Hc-iHmIl0I0;     SO51UQlxVyw; UI-ZteQ8JGU;
4Uwxr42JqYQ; 4yCUgx7H2zo;     Hp6wMUVb23c; hrsDBdnj5E8;     VAMzAIH12yc; Vlkrm8nLCJo;
5fHe1QFAHwQ; 5UBRUyofiiU;     ht4_hJBHejA; HtiY-X-wbIM;     VNM7Z7hir_I; vw61gCe2oqI;
7cXuWBMZfi8; 7GhY-ds2Kfk;     ItIb3nDGAPY; J1Yn84NVnSI;     wKbpqv-4Ku0; wwrreWsS3Vs;
8nN9lNJuqG4; 8pcWlyUu8U4;     JCwiW_YzSLM; jFhJjCmYilM;     xc1thxYM0Sk; XkWetbQHWlk;
8VcPF72MFsU; 9UcR9iKArd0;     jwuP_YHaySI; kCfwNcurP8o;     xV4YXmpcP3E; XztPtK7yAUk;
Ab8ds7NmigE; abPLDLV8O4s;     kh3ZwCkawiw; km9iS2tcRZE;     y36iVzlH4fs; YA1J-raGinQ;
aEi_4Cyx4Uw; AjUpiwvIa5A;     KmRBCgkiL3s; lkXFb1sMa38;     YWRa6Cz6m_M; zfvPPPc_UTs;
AledulxJKqo; azUbx8XwOlU;     moSFlvxnbgk; nbDPiaVEY7k;     zIEIvi2MuEk; ZoCyL_Pqzu8;
b7NLweylwYI; CaRlehRw9ig;     nbp3Ra3Yp74; nyc6RJEEe0U;     zQeygYqOn8g; ZXLO6mymzto;
D-3kFjFbSSE; D9BOTXFCpQA;     o8UCI7r1Aqw; oi7KPDi_yQI;
D9qlh0eAxG8; dc4duKuPrQ0;     p7iX3mTWESE; PB7xs7UpIfY;
dmz9Yb9dWck; dNEafGCf-kw;     pMWU8dEKwXw; PqQzjit7b1w;
e05BKmfKhTI; EEqZgGNXL7g;     q3bGY1jQ5Uw; QjKO10hKtYw;
```

**APPENDIX 2**

Table of frequencies for the categories of videos included in the case study sample

| Category | Frequency |
|---|---|
| Entertainment | 24 |
| Tech | 15 |
| Sports | 11 |
| Comedy | 9 |
| Education | 9 |
| News | 8 |

| Category | Frequency |
|---|---|
| Film | 7 |
| Animals | 4 |
| Music | 4 |
| People | 4 |
| Nonprofit | 3 |
| Howto | 1 |
| Travel | 1 |